\documentclass[11pt,aps,prd,onecolumn,nofootinbib,superscriptaddress]{revtex4-2}

\usepackage{amsmath,amssymb,amsfonts}
\usepackage{graphicx}
\usepackage{hyperref}
\usepackage{xcolor}
\usepackage{enumitem}

\newcommand{\mpl}{M_{\rm Pl}}
\newcommand{\gev}{\,\mathrm{GeV}}
\newcommand{\tev}{\,\mathrm{TeV}}

\newcommand{\Gtwo}{$G_2\,$}

\usepackage{xcolor}

\begin{document}

\title{Non-Perturbative Modulus Decay and Multi-Component Dark Matter}

\author{Gordon Kane}
\affiliation{Leinweber Institute for Theoretical Physics, University of Michigan, Ann Arbor, MI 48109, USA}
\affiliation{Physics Department, University of Michigan, Ann Arbor, MI 48109, USA}
\author{Leia Price}
\affiliation{Massachusetts Institute of Technology, Cambridge, MA 02139, USA}
\author{Luis Rufino}
\affiliation{Department of Physics, Syracuse University, Syracuse, NY 13244, USA}
\author{Scott Watson}
\affiliation{Department of Physics, Syracuse University, Syracuse, NY 13244, USA}
\author{Fred C. Adams}
\affiliation{Leinweber Institute for Theoretical Physics, University of Michigan, Ann Arbor, MI 48109, USA}
\affiliation{Physics Department, University of Michigan, Ann Arbor, MI 48109, USA}
\affiliation{Astronomy Department, University of Michigan, Ann Arbor, MI 48109, USA}

\date{\today}

\begin{abstract}
Non-thermal cosmological histories rest on the assumption that scalars displaced during inflation undergo coherent oscillations, dominate the energy density of the universe before Big Bang Nucleosynthesis, and decay perturbatively through gravitationally suppressed interactions.
In addition, usually a single dark matter component is assumed. This early matter-dominated era is the basis for the usual predictions of non-thermal WIMP production, axion dark matter, entropy generation, and dark radiation. This paper examines whether this picture is dynamically robust against non-perturbative decay of the modulus condensate. In this work, we study non-perturbative particle production from
oscillating moduli using the effective field theory appropriate to $G_2$ compactifications. We find that non-perturbative production of Wino-like fermions is strongly suppressed in the relevant parameter regime. In contrast, a modulus-dependent axion kinetic term admits narrow instability bands with growth rates that can exceed the Hubble rate. A linear Floquet analysis alone, however, cannot determine whether these bands significantly deplete the modulus condensate, since cosmic expansion, backreaction, rescattering, and higher-order operators in the effective theory can become important. As a result, the conventional modulus-dominated cosmology remains robust against Wino and gauge-field preheating, while the axion channel provides a potentially important modification that requires nonlinear study. If sufficiently efficient, axion production can alter the division of dark matter between Winos and axions and enhance the dark-radiation abundance. 
\end{abstract}

\maketitle
\newpage
\section{Introduction}
\label{sec:introduction}

Scalar moduli are a generic feature of string and M-theory compactifications.  They
parametrize the internal geometry of compactifications and must be stabilized in any phenomenologically viable vacuum
\cite{Coughlan1983,deCarlos1993,Banks1994,KaneSinhaWatson2015}.   During inflation,
Hubble-induced corrections generically shift their instantaneous minima, so that the
moduli emerge displaced from their late-time vacuum values
\cite{DineRandallThomas1995,Linde1996}.  When the Hubble scale falls to
$H\sim m_X$, a displaced modulus $X$ begins coherent oscillations about its minimum.
For an approximately quadratic potential the oscillation-averaged energy density
redshifts as nonrelativistic matter, $\rho_X\propto a^{-3}$, and can therefore come to
dominate over radiation before the modulus decays.  For gravitationally suppressed
couplings the sufficiently heavy moduli decay shortly before Big Bang nucleosynthesis
(BBN), reheating the visible sector and setting the initial conditions for the subsequent
radiation-dominated era \cite{Banks1994,Acharya2008Nonthermal,KaneSinhaWatson2015}.

The modulus-dominated era can dilute relics produced at earlier times, modify baryogenesis and dark-matter production, and generate dark radiation or other observable remnants of a non-standard pre-BBN expansion history \cite{Acharya2009WIMPMiracle,KaneSinhaWatson2015,AcharyaPongkitivanichkul2016}. In M-theory compactifications on manifolds of $G_2$ holonomy, these effects are particularly predictive.  Non-perturbative hidden-sector dynamics can stabilize the moduli and break supersymmetry, leading to the characteristic spectrum of the $G_2$-MSSM with heavy scalar superpartners and parametrically lighter gauginos \cite{Acharya2006Hierarchy,Acharya2007Moduli,Acharya2008G2MSSM}.  The lightest supersymmetric particle is generically Wino-like, and because the reheating temperature lies below the ordinary thermal freeze-out scale, its abundance is set by production from modulus decay followed by annihilation \cite{MoroiRandall2000,Acharya2008Nonthermal,Acharya2009WIMPMiracle}.  At the same time, axions arise intrinsically from the compactification.  The resulting axiverse can contain a QCD axion together with additional axion-like particles, giving a natural setting for multi-component dark matter in which Winos and axions coexist \cite{Acharya2010Axiverse,Acharya2012Generic}.  Entropy release from modulus decay can dilute axions that begin oscillating before reheating, whereas relativistic axions produced in modulus decays can contribute to $\Delta N_{\rm eff}$\footnote{Note that future experiments such as CMB-S4 will place tight constraints on $\Delta N_{\rm eff}$~\cite{CMB-S4:2026mge}.}.

The standard $G_2$-MSSM cosmology relies on the assumption that the coherently oscillating modulus condensate survives until it decays perturbatively.  However, oscillating scalar backgrounds can transfer energy non-perturbatively to lighter degrees of freedom through parametric resonance and tachyonic instabilities, processes that are familiar from the theory of preheating \cite{KofmanLindeStarobinsky1997,FelderEtAl2001}. Such mechanisms can arise through kinetic couplings, leading to non-perturbative gauge-field production \cite{DeskinsGiblinCaldwell2013} and axion production \cite{DealEtAl2025,LeedomEtAl2025}. 

In this paper, we test the robustness of the perturbative decay assumption within the effective field
theory appropriate to the $G_2$-MSSM.
A modulus-dependent axion kinetic function
$Z_a(X)(\partial a)^2$ periodically changes both the normalization and effective frequency of the axion modes.  The corresponding Floquet problem can contain narrow instability bands whose growth rate exceeds the Hubble rate for appropriate parameters, in agreement with the general expectation that moduli-dependent axion kinetic terms can support non-perturbative production \cite{DealEtAl2025,LeedomEtAl2025}. A large linear Floquet exponent, however, does not by itself imply substantial depletion  of the modulus condensate: The shrinking modulus amplitude, motion of comoving modes through the instability bands, backreaction, rescattering, and higher-order terms in the effective theory must all be included to determine the net energy transfer.  
The importance of the axion kinetic channel remains dependent on compactification-specific couplings and ultimately requires a nonlinear treatment. 

The remainder of this paper is organized as follows.  Section~II reviews the relevant ingredients of the $G_2$-MSSM, introduces a minimal modulus--Wino--axion effective description, and explores new nonperturbative effects from the oscillating modulus. Section~III discusses the implications for Wino and axion dark matter and for dark radiation.  We summarize the broader implications and future directions in Section~IV.

\section{Cosmological Ingredients of the \texorpdfstring{$G_2$}{G2}-MSSM}
\label{sec:g2}

In M-theory compactifications on \Gtwo holonomy manifolds, gauge, and matter sectors are localized on singular submanifolds, while the geometric moduli are stabilized by non-perturbative hidden-sector dynamics \cite{Acharya:2001gy,Acharya:2007rc,Acharya:2008zi,Acharya:2010zx,Acharya:2015zfk,KaneSinhaWatson2015}. In the resulting vacua, supersymmetry is broken by moduli $F$-terms and the gravitino mass naturally falls in the tens of TeV range.

\subsection{The Modulus-Dominated Era}
\label{sec:matterera}

Consider a modulus with a late-time minimum, and displacement $X$ from the minimum. During inflation the field is generically displaced from this minimum by Hubble-induced corrections to its potential. A schematic potential has the form 
\begin{equation}
  V(\psi) \sim \frac12 m_X^2 X^2
  - H_{\rm inf}^2 X^2
  + \frac{X^{4+2n}}{\mpl^{2n}} .
  \label{eq:modpot}
\end{equation}
The displacement can naturally be a substantial fraction of the Planck scale, so that 
\begin{equation}
  | X| 
  \sim \mpl \left(\frac{H_{\rm inf}}{\mpl}\right)^{1/(n+1)} .
\end{equation}
When $H \sim m_X$, the modulus becomes underdamped and begins coherent oscillations. The homogeneous mode satisfies the equation of motion\footnote{We will use the FLRW metric $ds^2=-dt^2+R^2(t) (dx^2+dy^2+dz^2)$}
\begin{equation}
  \ddot X + (3H+\Gamma_\psi)\dot X + \frac{\partial V}{\partial X}=0 .
\end{equation}
For an approximately quadratic potential, the oscillation-averaged equation of state becomes 
\begin{equation}
  \langle w_X\rangle \simeq 0,
\end{equation}
so that
\begin{equation}
  \rho_X \propto R^{-3}, \qquad {\rm and} \qquad 
  \rho_r \propto R^{-4}.
\end{equation}
Even if radiation dominates initially, the modulus energy density thus grows relative to radiation and generically comes to dominate before it decays.
We will therefore approximate
\begin{align}
    H^2\approx \frac{m_X^2X_0^2}{6M_{\rm pl}^2}
\end{align}
where $X_0$ is the amplitude of oscillation for the modulus. 

This matter-dominated era is the origin of the non-thermal dark matter scenario. Modulus decay injects entropy into the background universe and produces LSPs $\chi$ at a temperature below thermal freeze-out. If too many LSPs are produced, annihilations reduce their number density until a steady-state is reached where 
\begin{equation}
  n_\chi \langle \sigma v \rangle \sim H(T_{\rm rh}) .
\end{equation}
The final abundance is controlled mainly by $T_{\rm rh}$, the LSP mass, and the annihilation cross section. A useful parametric form is
\begin{equation}
\Omega_\chi h^2 \sim 0.1
\left(\frac{m_\chi}{100\gev}\right)^3
\left(\frac{10.75}{g_*(T_{\rm rh})}\right)^{1/4}
\left(\frac{3\times 10^{-7}\gev^{-2}}{\langle\sigma v\rangle}\right)
\left(\frac{100\tev}{m_X}\right)^{3/2},
\label{eq:wimpabund}
\end{equation}
up to order-one branching and decay coefficients.

The same modulus decay also dilutes axion abundances and any dangerous relics produced at earlier times. 
Several other features of this scenario are important for cosmology:
Firstly, the moduli must be heavy enough to decay before BBN \cite{Banks:1995dt,Dine:1995uk,deCarlos:1993jw,KaneSinhaWatson2015}. Their decay widths are gravitationally suppressed, schematically
\begin{equation}
  \Gamma_X \simeq D_X \frac{m_X^3}{\mpl^2},
  \label{eq:width}
\end{equation}
where $D_X$ is an order-unity coefficient.
The corresponding reheating temperature is given by 
\begin{equation}
  T_{\rm rh} \simeq \left( \Gamma_X \mpl \right)^{1/2}
  \left(\frac{90}{\pi^2 g_*(T_{\rm rh})}\right)^{1/4} .
  \label{eq:Trh}
\end{equation}
For moduli mass in the range of tens to hundreds of TeV, the reheating temperature naturally falls above the BBN bound, but well below the usual thermal freeze-out temperature for electroweak dark matter.

Secondly, the low-energy spectrum contains heavy scalars and lighter gauginos. The lightest supersymmetric particle is typically Wino-like, with a mass near the electroweak scale. Since the reheating temperature is low, the Wino abundance is not set by ordinary thermal freeze-out. Instead, moduli decay produces LSPs non-thermally, after which annihilations reduce the abundance to a quasi-fixed-point value.

Third, the compactification contains a collection of axions associated with
the moduli sector.  Writing the chiral moduli as
\begin{equation}
 z_j=t_j+i s_j,
\end{equation}
the fields $s_i$ describe geometric deformations of the $G_2$ manifold, whereas the periodic fields $t_i$ descend from the eleven-dimensional three-form.  The physical saxions and axions are in general different linear combinations of these fields, 
\begin{equation}
 x_I=U^{(s)}_{Ij}\,s_j\ \ ,
 \qquad
 a_I=U^{(t)}_{Ij}\,t_j,
\end{equation}
obtained after diagonalizing the kinetic and mass matrices.  Consequently, a single late-decaying modulus need not be paired with a single light axion in the mass eigenstate basis.

In the fluxless $G_2$ vacua relevant to the $G_2$-MSSM, the leading hidden-sector non-perturbative terms stabilize the geometric moduli and one axionic combination.  The remaining axionic directions are lifted by subleading non-perturbative effects, such as membrane instantons or additional gaugino condensates.  Their masses are thus exponentially hierarchical rather than tied directly to the modulus mass scale, giving an $M$-theory realization of an axiverse.  One linear combination can couple to QCD and remain light enough to implement the Peccei--Quinn mechanism, while the other combinations behave as axion-like particles.  Typical decay constants are near the
unification scale, although their precise values and anomaly coefficients are compactification dependent \cite{Acharya:2010zx}. 

The cosmological role of these fields depends on when they begin to oscillate relative to modulus decay.  Axions with $H\sim m_{a_I}$ during the modulus-dominated era acquire a misalignment abundance that is subsequently diluted by the entropy released at reheating.  Axions that begin to oscillate after reheating instead follow the standard radiation-dominated misalignment history.  In addition, perturbative modulus decays can produce relativistic axions, while derivative modulus--axion couplings can generate axions non-perturbatively.  The former contributes to dark radiation through the branching ratio $B_{X\to aa}$, whereas the latter is controlled by independent coefficients in the axion kinetic function and must be checked separately from gauge-field production \cite{Acharya:2015zfk,DealEtAl2025}.

\subsection{A minimal modulus-Wino-axion effective theory}

A model Lagrangian can be written in the form 
\begin{align}
\frac{{\cal L}_{\rm eff}}{\sqrt{-g}}
={}& \frac{M_{\rm Pl}^2}{2}R_4
-\frac12(\partial X)^2-V_X(X)
-\frac12 Z_a(X)(\partial a)^2-V_a(a)
\nonumber\\
&-\frac14\sum_{r=1}^{3} Z_r(X)
 F^{(r)}_{\mu\nu}F_{(r)}^{\mu\nu}
+\sum_{r=1}^{3}\frac{\alpha_r C_r}{8\pi}
 \frac{a}{f_a}F^{(r)}_{\mu\nu}\widetilde F_{(r)}^{\mu\nu}
\nonumber\\
&+\frac12\,\overline{\widetilde W}^{A}
 \left(i\gamma^\mu D_\mu-m_{\chi\rm,eff}(X)\right)\widetilde W^{A}
+\frac{C_{a\widetilde W}}{2f_a}\,
 \partial_\mu a\,
 \overline{\widetilde W}^{A}\gamma^\mu\gamma^5\widetilde W^{A}
+{\cal L}_{\rm visible}+\cdots .
\label{eq:total-toy-EFT}
\end{align}
Here $A=1,2,3$ labels the adjoint of $SU(2)_L$, and the neutral component of $\widetilde W^A$ becomes the Wino-like LSP after electroweak symmetry breaking. The covariant derivative contains the ordinary electroweak interactions that determine the dominant Wino annihilation and coannihilation rates.  A Wino-like neutralino rather than a pure Wino can be incorporated by replacing $\widetilde W^3$ with the appropriate neutralino mixing eigenstate. 
The coefficient $C_{a\widetilde W}$ is optional; it represents a direct derivative axion-Wino interaction allowed in the low-energy axion-like particles (ALPs) EFT, but it may vanish or be loop suppressed in a minimal realization where the dominant axion coupling is through the gauge anomaly.  More generally, the $X$-dependent couplings are not independent in a UV completion.  They are determined by the $G_2$ moduli-space metric, the visible-sector gauge kinetic functions, the axion and modulus mixing matrices, and the supersymmetry-breaking auxiliary fields. 


The functions appearing in Eq.~\eqref{eq:total-toy-EFT} may be expanded about
the late-time vacuum as
\begin{align}
 Z_a(X)&=1+\kappa_a\frac{X}{\Lambda_a}
 +\widetilde\kappa_a\frac{X^2}{\Lambda_a^2}+\cdots,
\nonumber\\
 Z_r(X)&=1+c_r\frac{X}{\Lambda_r}+\cdots,
\nonumber\\
 m_{\chi,\rm eff}(X)&=m_\chi\left(1+c_\chi\frac{X}{\Lambda_\chi}+\cdots\right),
\label{eq:EFT-expansions}
\end{align}
with
\begin{equation}
 V_X(X)=\frac12m_X^2X^2+\cdots
 \qquad {\rm and} \qquad
 V_a(a)=m_a^2f_a^2
 \left[1-\cos\left(\frac{a}{f_a}\right)\right].
\end{equation}

This form makes the distinct production channels transparent.  Expanding $Z_r(X)$ gives the CP-even operator $X F^2/\Lambda_r$ used in a gauge-field resonance analysis, which was already performed in Ref.~\cite{Giblin:2017wlo} and was found to be inefficient.  Expanding $Z_a(X)$ gives the derivative interaction
\begin{equation}
 {\cal L}_{Xaa}=-\frac{\kappa_a}{2\Lambda_a}
 X(\partial a)^2+\cdots,
\end{equation}
which controls perturbative decay $X\to aa$ and can also generate tachyonic or
parametric axion production.  Expanding $m_{\chi,\rm eff}(X)$ gives
\begin{equation}
 {\cal L}_{X\widetilde W\widetilde W}
 =-\frac{c_\chi m_\chi}{2\Lambda_\chi}
 X\overline{\widetilde W}^{A}\widetilde W^{A}+\cdots,
\end{equation}
which schematically represents modulus decay into Winos (where $\chi$ in the coefficients represents the Wino).  Finally, the $aF\widetilde F$ terms encode the QCD axion coupling and possible ALP couplings to the electroweak and QCD gauge sectors.

\subsection{A Minimal Model for Axion and Wino Dark Matter}
\label{subsec:multicomponent-simulation}

The analysis that follows considers whether non-perturbative particle production can significantly deplete the coherent modulus condensate before its perturbative decay. In the ($G_2$)-MSSM, the answer to this question extend directly to the dark sector. The late decay of the modulus both determines the abundance of a Wino-like lightest supersymmetric particle and dilutes, produces, or otherwise modifies the abundance of axions. A consistent working model should thus include both components of the resulting multi-component dark matter scenario.

The two dark matter candidates are governed by physically different processes. Axions are bosonic fields and can be amplified directly by the coherent oscillations of the modulus. In particular, a modulus-dependent axion kinetic term produces a periodically varying normalization of the axion field and may lead to parametric or tachyonic amplification. The Wino-like LSP, by contrast, is a fermionic particle whose late-time abundance is primarily determined by perturbative modulus decay followed by Wino pair annihilation. It is therefore useful to adopt a hybrid description in which the modulus and axion are evolved as fields, while the Wino abundance is evolved using a Boltzmann equation.

\begin{figure}
    \centering
    \includegraphics[width=0.8\linewidth]{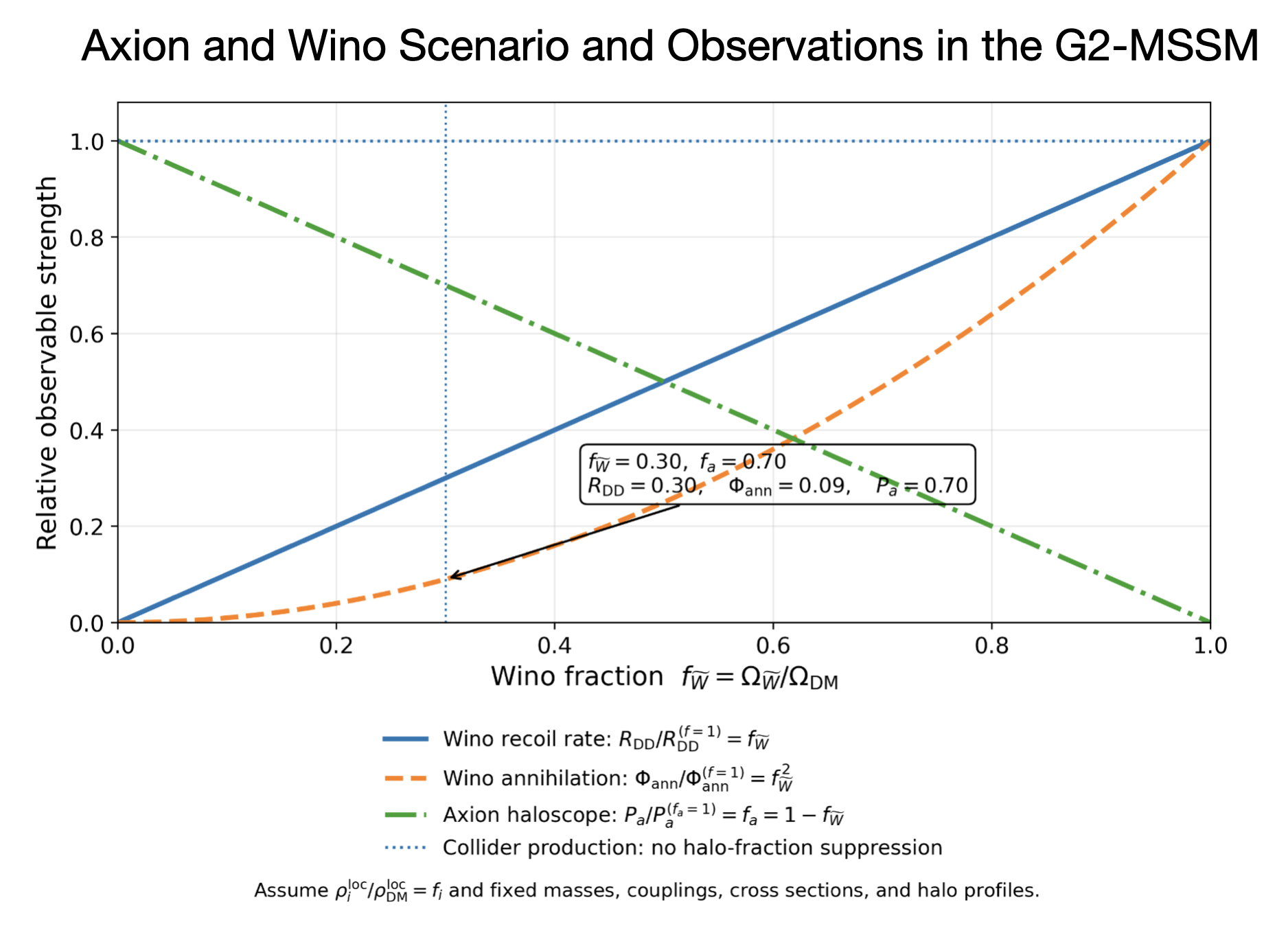}
    \caption{Collider and Cosmological observations as the percent of the amount of Wino dark matter.}
    \label{fig:wino}
\end{figure}

\begin{figure}
    \centering
    \includegraphics[width=1.0\linewidth]{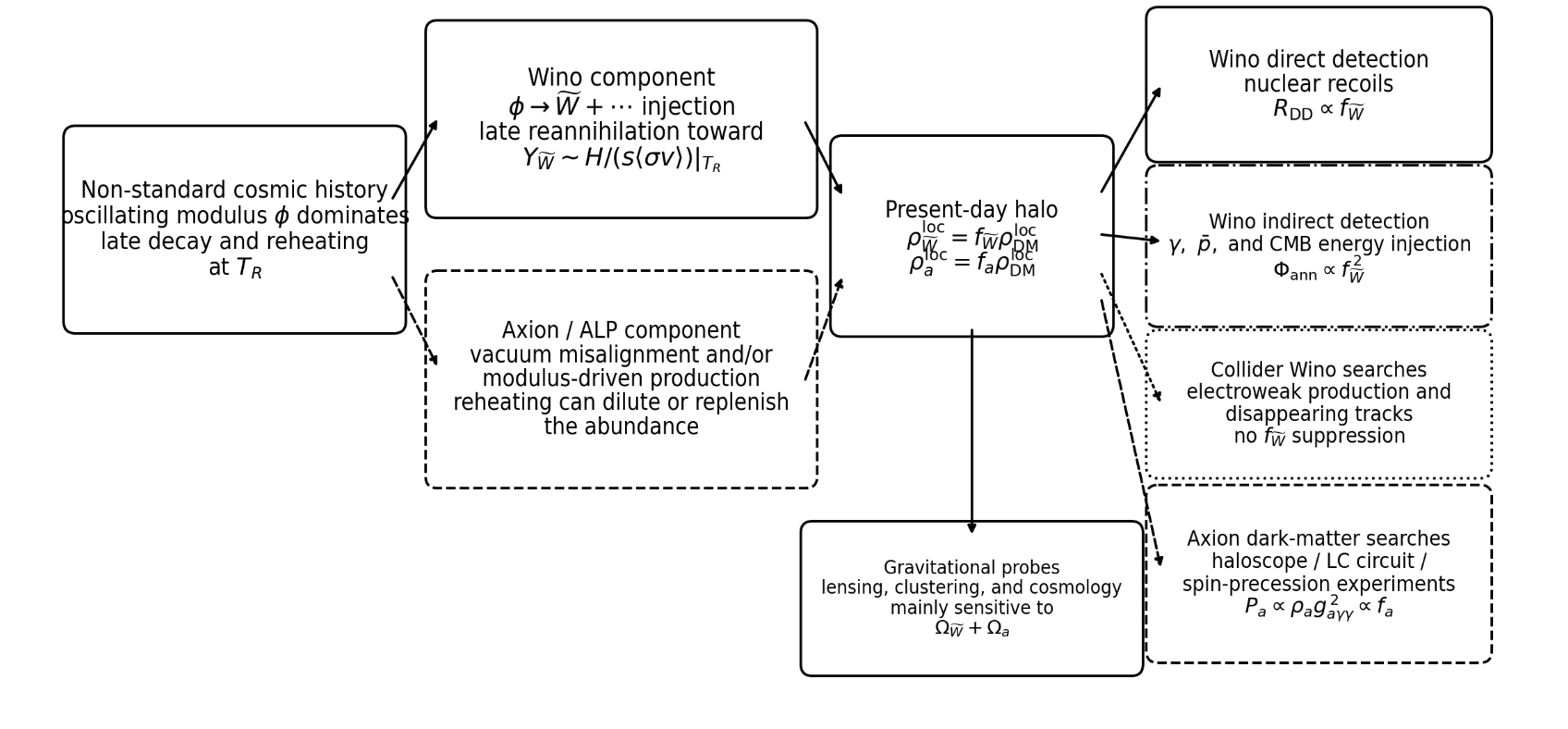}
    \caption{A flowchart of the connection between collider and cosmological observations for the scenario of two dark matter candidates where $\Omega_{DM}=\Omega_{\tilde W}+\Omega_a$ and $f_{\tilde W}+f_a=1$.}
    \label{fig:axion}
\end{figure}

A minimal effective Lagrangian resulting from the simplifying of Eq.~\ref{eq:total-toy-EFT} is
\begin{equation}
\begin{aligned}
\frac{{\cal L}}{\sqrt{-g}}
&=
-\frac{1}{2}(\partial X)^2
-\frac{1}{2}
Z_a(X)(\partial a)^2
-\frac{1}{2}m_X^2X^2
-\frac{1}{2}m_a^2a^2
\\&\quad
+
\frac{1}{2}\overline{\chi}
\left(
i\gamma^\mu D*\mu-m_\chi
\right)\chi
-\frac{c_\chi m_\chi}{2\Lambda_\chi}
X\overline{\chi}\chi ,
\end{aligned}
\label{eq:multicomponent-toy-lagrangian}
\end{equation}
where
\begin{equation}
Z_a(X)=
1+\kappa_a\frac{X}{\Lambda_a}.
\label{eq:linear-axion-kinetic-function}
\end{equation}
For the Floquet analysis below we truncate $Z_a(X)$ at linear
order in $X/\Lambda_a$ in order to isolate the leading derivative
interaction. The quadratic and higher-order terms appearing in
Eq.~(14) are neglected at this stage. This truncation is appropriate
when $X/\Lambda_a$ is sufficiently small, or when the corresponding
higher-order coefficients are suppressed. As the system approaches
the boundary of the EFT regime, however, terms such as
$\widetilde{\kappa}_a X^2/\Lambda_a^2$ should be restored.
Here $\Lambda_a$ and $\Lambda_\chi$ denote the scales suppressing the modulus couplings to the axion and Wino sectors, respectively. The dimensionless parameters $\kappa_a$ and $c_\chi$ contain the compactification-dependent information associated with the relevant modulus eigenstate. The field $\chi$ should be interpreted as an effective Wino-like LSP rather than as a complete description of the neutralino and chargino sectors.

The effective theory must remain under perturbative control; in particular, the modulus-dependent axion normalization must remain positive, 
\begin{equation}
Z_a(X)>0,
\label{eq:positive-kinetic-function}
\end{equation}
and the dimensionless expansion parameters should satisfy the constraints 
\begin{equation}
\left|
\kappa_a\frac{X}{\Lambda_a}
\right|
\lesssim 1 
\qquad {\rm and} \qquad 
\left|
c_\chi\frac{X}{\Lambda_\chi}
\right|
\lesssim 1.
\label{eq:multicomponent-eft-control}
\end{equation}
These conditions prevent the model from being extrapolated into a regime in which neglected higher-dimensional operators become equally important.

The axion potential has been approximated as quadratic. This form represents the small-field limit of the usual periodic potential,
\begin{equation}
m_a^2f_a^2
\left[
1-\cos\left(\frac{a}{f_a}\right)
\right]
\simeq
\frac{1}{2}m_a^2a^2,
\qquad
\left|\frac{a}{f_a}\right|\ll 1.
\label{eq:axion-small-field-expansion}
\end{equation}
The quadratic approximation is sufficient for studying the initial amplification of axion fluctuations. The full cosine potential can subsequently be restored when investigating order-unity misalignment angles, axion self-interactions, or nonlinear field evolution.
To zeroth order, the axion can be treated as effectively massless over a modulus oscillation,
\begin{equation}
m_a\ll m_X,
\end{equation}
so that $m_a$ may be set to zero in our baseline Floquet analysis. The modulus-dependent kinetic term isolates the effects associated with the axion potential.

The modulus is a homogeneous condensate in a quadratic potential,
\begin{equation}
X(t)\simeq X_0(t)\cos(m_Xt),
\label{eq:modulus-background-multicomponent}
\end{equation}
where $X_0(t)$ decreases as a result of Hubble expansion.
This oscillating modulus makes $Z_a$ time dependent. 

Taking all these ingredients together, we have an EOM for the axion modes in terms of the scale factor $R(t)$,
\begin{equation}
\ddot a_k
+
\left(
3H+\frac{\dot Z_a}{Z_a}
\right)\dot a_k
+
\left[
\frac{k^2}{R^2}
+
\frac{m_a^2}{Z_a}
\right]a_k
=0.
\label{eq:axion-mode-equation-simulation}
\end{equation}
The effective damping term and the axion frequency vary periodically. Equation~\eqref{eq:axion-mode-equation-simulation} therefore permits instability bands whose strength and duration depend on the ratio $\kappa_a X_0/\Lambda_a$, the mass ratio $m_a/m_X$, the comoving momentum $k$, and the expansion rate.

\begin{figure}
    \centering
    \includegraphics[width=0.49\linewidth]{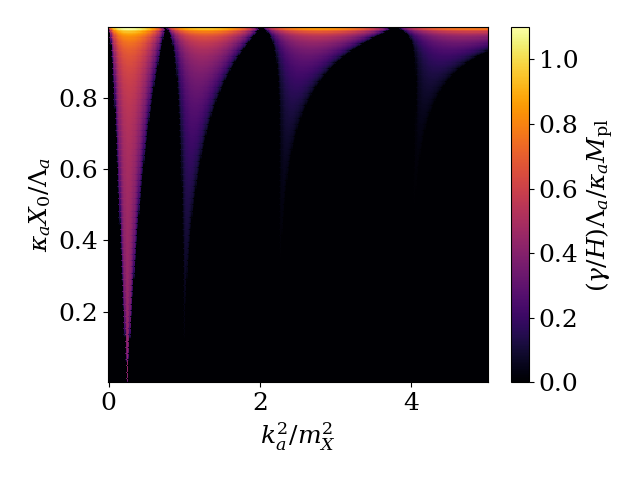}
    \includegraphics[width=0.49\linewidth]{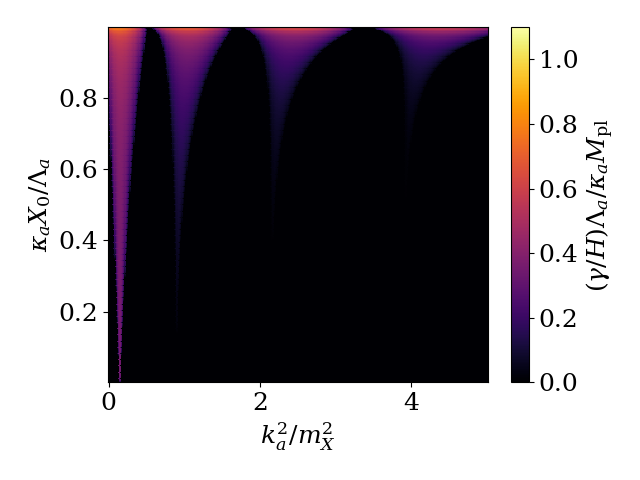}
    \caption{Left: Floquet map for the growth of axion modes $A_k$ from Eqn.~(\ref{eqn:axion-eom}) in the limit where  $m_a/m_X\to0$ and $H/m_X\to0$. We consider expansion by displaying $\gamma/H$. Axion production would be considered efficient if $\gamma/H\gtrsim1$, which depends on the values of $\kappa_a$ and $\Lambda_a$. Since $\Lambda/M_{\rm pl}\sim10^{-3}$, and $\kappa_a\sim1$ these Floquet exponents will be of order $\gamma/H\sim10^2$, meaning that resonance is very efficient within these admittedly narrow instability bands. The most efficient resonance will occur for $k_a/m_X\approx 0.2$.
    Right: Left: Floquet map for the growth of axion modes $A_k$ from Eqn.~\ref{eqn:axion-eom} with $m_a/m_X=0.1$ and $H/m_X\to0$. The instability bands are dampened and slightly shifted compared to the massless axion case.}
    \label{fig:axion-floquet}
\end{figure}

We scale the axion field by
\begin{align}
    A_k\equiv \sqrt{R^3 Z_a}\,a_k
    \label{eqn:A-a-rescale}
\end{align}
to get an EOM for $A_k$ without a friction term.
This step allows us to perform a more accurate Floquet analysis, which is typically calculated in the absence of friction to isolate the effect of the periodic effective mass.
Although ${H}/{m_X} \ll 1$ permits the Hubble friction to be dropped for this purpose, the ${\dot{Z_a}}/{Z_a}$ term in Eqn ~\ref{eq:axion-mode-equation-simulation} is itself a significant source of non-perturbative axion production and must be retained.
This EOM with the field redefinition
is
\begin{align}
    \ddot A_k + \left(\frac{k^2}{R^2} + \frac{m_a^2}{Z_a}
    - \frac{9}{4}H^2 -\frac{3}{2}\dot H - \frac{3\dot Z_a}{2Z_a} H+\frac{\dot Z_a^2}{4Z_a^2} -\frac{\ddot Z_a}{2Z_a} \right)A_k = 0\ .
    \label{eqn:axion-eom}
\end{align}
The Floquet analysis of this EOM is shown in Fig.~\ref{fig:axion-floquet}, where we display Floquet exponents $\gamma_k$ (also termed ``growth rates'') defined by Floquet's theorem,
\begin{align}
    A_k(t) = e^{\mu_k t}P_1(t)+e^{-\mu_kt} P_2(t)\ \ , \quad \gamma_k\equiv{\rm Re}\,\mu_k
\end{align}
where $P_{1,2}$ have periodicity matching the modulus' period.
The resonance bands are narrow, meaning most of the parameter space has $\gamma=0$.
However, within those narrow bands the resonance efficiency is very high, $\gamma/H\sim10-100$, meaning that axions can be exponentially produced very efficiently.
While we cannot estimate an exact resulting axion abundance, the possibility that axions are produced in excess in contrast with the standard \Gtwo-MSSM picture is an interesting possible consequence of considering the coherently oscillating modulus. 

In \cite{Greene:2000ew}, Greene and Kofman developed an analytic, nonperturbative theory of fermionic preheating from a Yukawa-coupled, oscillating inflaton, using successive parabolic scatterings to track fermion occupation numbers. They show that production is rapid but Pauli-limited; cosmic expansion makes it stochastic, filling a momentum-space “Fermi sphere” of radius \(\sim q^{1/4}m_\phi\), while even superheavy fermions can be efficiently produced whenever their effective mass \(m_\psi+h\phi\) crosses zero.  
This is important for understanding how Wino production can occur from the moduli (analogous to the inflaton story).

From the modulus--Wino interaction in Eq.~(17),
\begin{equation}
{\cal L}_{X\chi\chi}
=
-\frac{c_\chi m_\chi}{2\Lambda_\chi}
X\,\bar{\chi}\chi ,
\end{equation}
the oscillating modulus background gives rise to an effective Wino mass
\begin{equation}
m_{\chi,{\rm eff}}(t)
=
m_\chi
\left(
1+c_\chi\frac{X(t)}{\Lambda_\chi}
\right),
\label{eq:wino_effective_mass}
\end{equation}
up to an overall sign convention for the interaction.

After an appropriate rescaling of the fermion helicity eigenmode functions to remove the
Hubble dilution, the second-order momentum mode equation is
\begin{equation}
\ddot{\Psi}_k
+
\left[
\frac{k^2}{R^2}
+
m_{\chi,{\rm eff}}^2
-
i\,\frac{1}{R}
\frac{d}{dt}
\left(
R\,m_{\chi,{\rm eff}}
\right)
+
\Delta(R)
\right]
\Psi_k
=0 ,
\label{eq:fermion_mode}
\end{equation}
where
\begin{equation}
\Delta(R)
=
\frac{1}{4}H^2
-
\frac{1}{2}\frac{\ddot a}{a}.
\end{equation}
We will neglect $\Delta$ to find the growth of modes during one modulus oscillation period, then compare that growth rate to $H$ to determine the resonance efficiency.
The instantaneous frequency is therefore
\begin{equation}
\omega_k^2(t)
=
\frac{k^2}{R^2}
+
m_{\chi,{\rm eff}}^2(t).
\end{equation}

\begin{figure}
    \centering
    \includegraphics[width=0.49\linewidth]{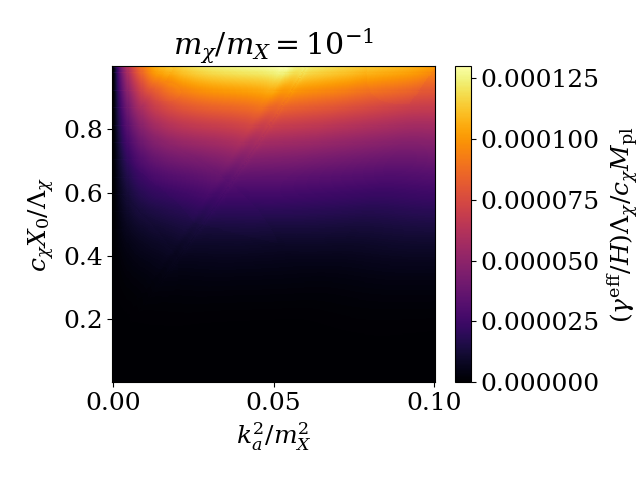}
    \includegraphics[width=0.49\linewidth]{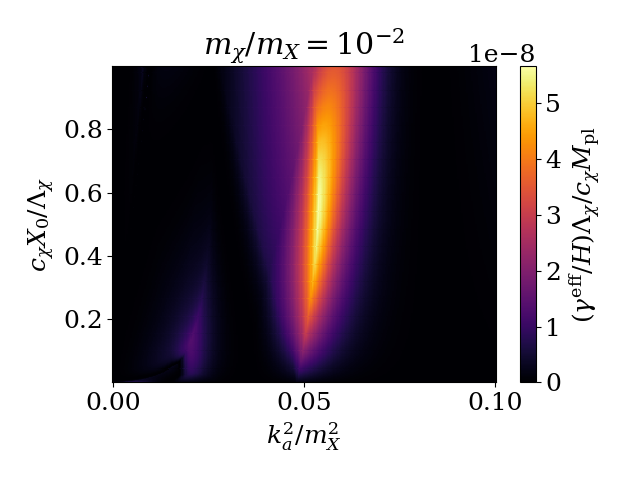}
    \includegraphics[width=0.49\linewidth]{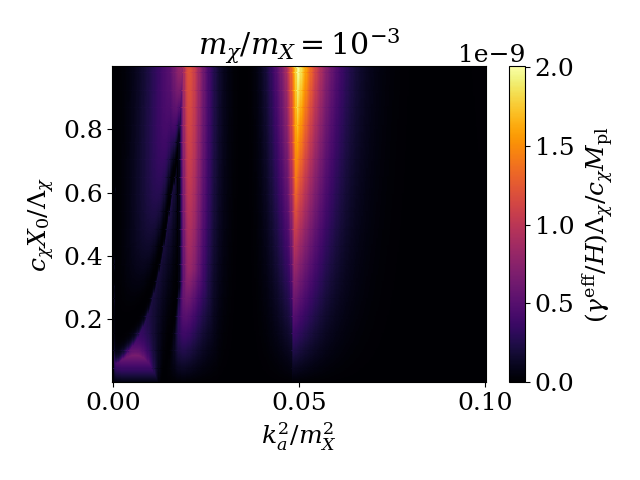}
    \includegraphics[width=0.49\linewidth]{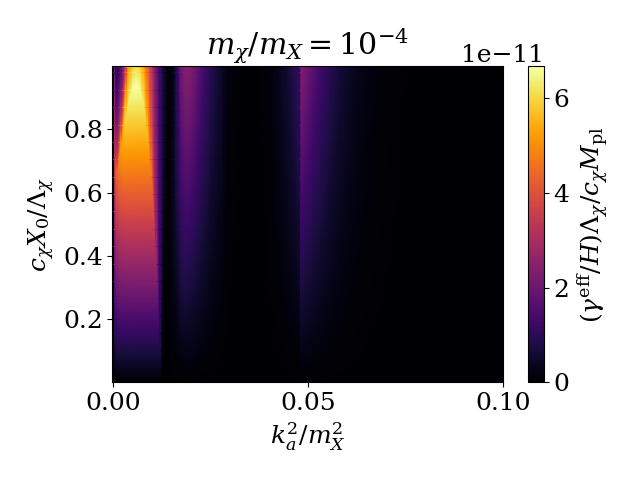}
    \caption{Fermion stability maps with a growth rate $\gamma^{\rm eff}$ defined by eqn.~\ref{eqn:gamma-eff} characterizing the growth of fermion modes for four different ratio choices $m_\chi/m_X$. The Hubble rate $H$ is rightly approximated to be dominated by the modulus. These Floquet charts are interesting but ultimately show that resonant production of Winos is inefficient in \Gtwo-MSSM.}
    \label{fig:fermion-floquet-maps}
\end{figure}

The number density of a Wino mode after a single oscillation period $T=2\pi/m_X$ of the modulus is (Greene-Kofman's eqn. (19))
\begin{align}
    n_k(T)=\frac{k^2}{2\omega_k^2}\left({\rm Im}\big(\Psi^{(1)}_k\big)\right)^2 
\end{align}
where the $\Psi^{(1)}$ is the solution with initial conditions $\Psi^{(1)}(0)=1$, $\dot \Psi^{(1)}(0)=0$.
In analogy with Floquet's theorem and the growth of axionic and bosonic modes, we can define an effective growth rate $\gamma^{\rm eff}$ through the number density per fermion mode,
\begin{align}
    n_k(T)\equiv e^{2\gamma_k^{\rm eff}T}-1
    \label{eqn:gamma-eff}
\end{align}
which holds as long as $n_k<1$ to respect Pauli blocking.
This expression gives us the ability to make Floquet charts to determine the effectiveness of resonance in producing Winos.
These effective growth rates are compared to the cosmic expansion rate in Fig.~\ref{fig:fermion-floquet-maps}.
The growth rates are indeed very low such that fermion occupation number is safely below unity.
The \Gtwo-MSSM gives very little resonant fermion production because of the low expected ratio $m_\chi/m_X$, which controls the relative magnitude of the oscillating terms in the fermion EOM.
However, Winos with low masses are more kinematically available to be produced thermally.

Let $n_\chi$ denote the total number density of Wino-like LSPs. Its evolution is described schematically by
\begin{equation}
\dot n_\chi + 3H n_\chi = N_\chi {\rm Br}_{\chi} \Gamma_X \frac{\rho_X}{m_X} - \langle\sigma v\rangle_\chi \left(n_\chi^2 - n_{\chi,{\rm eq}}^2\right)
\label{number}
\end{equation}
 For a direct two-body decay ($X\rightarrow\chi\chi$), one has ($N_\chi=2$). The final term accounts for Wino annihilation, with ($\langle\sigma v\rangle_\chi$) denoting the thermally averaged annihilation cross section.

Because the modulus reheats the universe below the conventional Wino freeze-out temperature, the equilibrium abundance is generally negligible during the principal production epoch. Then \eqref{number} reduces to
\begin{equation}
\dot n_\chi+3Hn_\chi
\simeq
N_\chi {\rm Br}_{\chi}
\Gamma_X\frac{\rho_X}{m_X}-
\langle\sigma v\rangle_\chi n_\chi^2 
\label{eq:nonthermal-wino-equation}
\end{equation}
The first term produces Winos through modulus decay, while the second reduces an initially overproduced abundance through pair annihilation. When annihilations are efficient, the Wino density approaches the familiar nonthermal quasi-fixed-point behavior,
\begin{equation}
n_\chi\langle\sigma v\rangle_\chi
\sim H.
\label{eq:wino-quasifixed-point}
\end{equation}

The remaining question, answerable by future work with lattice simulations, is whether non-perturbative axion production removes a significant fraction of the modulus energy before ($H\simeq\Gamma_X$). If it does not, the standard modulus-decay calculation of the nonthermal Wino abundance remains intact. If it does, the reduced modulus density, altered reheating history, and additional axion population must all be included when determining the final division of dark matter between Winos and axions.

\section{Consequences for Multi-Component Dark Matter}
\label{sec:dm}


\subsection{Axions}

The resonant production of axions has competing factors determining its efficiency.
The relevant parameter space of the Floquet chart admits narrow resonance bands.
On the other hand the Floquet exponents within those narrow bands are high, $\gamma/H\sim 10-100$, so that axion production in the \Gtwo-MSSM may be much greater than predicted by the perturbative estimate. 

This possibility has several consequences for the standard nonthermal cosmology associated with $G_2$ compactifications. First, rapid transfer of energy from the modulus condensate into axions can shorten or otherwise modify the modulus-dominated era. Since the conventional $G_2$-MSSM scenario relies on late modulus decay to reheat the visible sector and generate nonthermal Wino dark matter, an early depletion of the modulus condensate can alter both the duration of early matter domination and the resulting Wino abundance.

Second, efficient nonperturbative production can generate a large population of relativistic axions. For sufficiently light axions, the characteristic momentum at production is of order $p_a \sim {\cal O}(m_X)$, so these particles initially contribute as dark radiation rather than as cold axion dark matter. Their abundance is therefore constrained by limits on additional relativistic degrees of freedom~\cite{DealEtAl2025}.
There is also the possibility of the formation of axion miniclusters~\cite{Nelson_2018} or oscillons~\cite{Amin_2012}.

Third, axion production and Wino production are no longer independent processes. If a fraction of the modulus energy is transferred into axions before perturbative decay, both the entropy released into the visible sector and the number of supersymmetric particles produced by the subsequent modulus decay are modified. The resulting cosmology naturally contains coupled axion and Wino relic populations.

\subsection{Wino-like LSPs}

In the \Gtwo-MSSM the LSP is generically Wino-like. Because the reheating temperature from modulus decay is below the ordinary freeze-out temperature, the Wino abundance is non-thermal. The final density is determined by the competition between production from modulus decay and subsequent annihilation. If nonperturbative effects do not destroy the condensate, the standard calculation based on nonthermal production remains applicable.

This point is important because the Wino abundance is sensitive to the reheating temperature. A premature depletion of the modulus condensate would raise or lower the effective reheating history and modify the final relic density. Our result shows that, within the controlled EFT, this may not occur, and further numerical study is needed to find more accurate particle abundances resulting from this scenario.

\section{Conclusions}
\label{sec:conclusions}

The primary motivation of this paper was to explore a fundamental theory that makes observational predictions for more than one dark matter candidate and determine the extent to which parametric resonance plays a role.
The standard non-thermal cosmology of the $G_2$-MSSM results in the survival of a coherently oscillating modulus condensate until its perturbative decay, which must occur prior to BBN. The condensate generates an early matter-dominated era, reheats the visible sector shortly before BBN, and sets the initial conditions for non-thermal Wino production, axion dark matter, entropy generation, and dark radiation. In this work we have examined whether non-perturbative particle production can invalidate this basic cosmological picture.

Our linear Floquet analysis of axions exhibits narrow but strong
instability bands in which the growth rate can exceed the Hubble rate,
particularly for axions that are lighter than the modulus. A large
Floquet exponent, however, does not by itself establish substantial
depletion of the modulus condensate. Determining the net energy
transferred to axions requires following the shrinking modulus
amplitude, the motion of comoving modes through the instability bands,
backreaction, rescattering, and the nonlinear evolution of the axion
sector.

The Floquet analysis presented here should therefore be regarded as a
leading-order benchmark based on the truncation
$Z_a(X)=1+\kappa_a X/\Lambda_a$. As discussed following Eq.~(19),
quadratic and higher-order terms are neglected in order to isolate the
leading derivative interaction. Near the boundary of the EFT regime,
however, terms such as
$\widetilde{\kappa}_a X^2/\Lambda_a^2$ need not be negligible. These
terms can introduce additional harmonics into the periodically varying
axion normalization and would introduce much more parameter space to the Floquet analysis. A quantitative calculation of condensate depletion
should therefore ultimately include these operators together with
backreaction and rescattering.

If nonperturbative production leads to an axion abundance
significantly larger than in the conventional $G_2$-MSSM cosmology,
the consequences extend beyond simply increasing the axion fraction
of dark matter. The resonantly produced axions have characteristic
momenta of order a fraction of $m_X$ and are therefore initially
relativistic for the light axions considered here. Their first
cosmological manifestation would consequently be an enhanced
dark-radiation component and potentially a larger contribution to
$\Delta N_{\rm eff}$. Depending on their masses and subsequent
redshifting, some of this population may later become nonrelativistic
and contribute to the axion dark-matter abundance.

At the same time, energy transferred from the modulus into axions is
no longer available for visible-sector reheating or for the
perturbative production of Wino-like LSPs. Efficient axion production
can therefore modify the reheating history, the entropy dilution of
the misalignment population, and the final Wino abundance
simultaneously. The resulting dark sector should thus be viewed as a
coupled two-component system rather than as two independently
calculated relic abundances.

Enhanced axion production would also modify the observational
complementarity illustrated in Figs.~1 and~2. A smaller Wino fraction
suppresses direct- and indirect-detection signals associated with the
Wino component, while an increased axion fraction enhances the
relative importance of axion searches. If the axion fluctuations
become nonlinear, additional signatures associated with small-scale
structure, axion miniclusters, or localized axion configurations may
also become relevant. Thus, determining the integrated efficiency of
the resonance is essential not only for establishing whether the
modulus-dominated era survives, but also for predicting how the
dark-matter abundance is divided between Winos and axions.

Provided that nonperturbative axion production remains weak and
extracts only a subdominant fraction of the modulus energy, the usual
nonthermal predictions of the $G_2$-MSSM remain intact. Perturbative
modulus decay sets the reheating temperature, while the abundance of
the Wino-like LSP is determined by the competition between decay
production and subsequent annihilation. Axions beginning their
oscillations before reheating are diluted by the entropy released in
modulus decay, whereas relativistic axions produced perturbatively
contribute to dark radiation according to the appropriate branching
fractions. Conversely, if the axion instability becomes nonlinear,
the resulting change in the reheating history and the additional axion
population must be included when determining the final division of
dark matter between Winos and axions.

\begin{acknowledgments}
We thank Bobby Acharya for useful discussions.
This research was supported in part by DOE grant DE-FG02-85ER40237 and the Simons Center. The research at the University of Michigan was supported in part by the Leinweber Institute for Theoretical Physics. 
\end{acknowledgments}

\appendix
\section{Fermionic Hill's equation}

Here we include a short note on solving the fermionic equation of motion.
The real and imaginary parts of $\Psi_k$ are denoted by $R_k\equiv {\rm Re}\Psi_k$, $I_k\equiv {\rm Im}\Psi_k$, and their EOMs are
\begin{align}
    \ddot R_k+\omega_k^2R_k+\dot m_{\chi,\rm eff}I_k=0 \\
    \ddot I_k + \omega_k^2I_k-\dot m_{\chi,\rm eff}R_k=0\ .
\end{align}
The system can be diagonalized by introducing the complex combinations
\begin{equation}
    Z_{k+} = R_k + i I_k,
    \qquad
    Z_{k-} = R_k - i I_k .
\end{equation}
The corresponding equations of motion are
\begin{equation}
    \ddot{Z}_{k\pm}
    +
    \left(
        \omega_k^2
        \mp i\dot{m}_{\chi,\mathrm{eff}}
    \right)
    Z_{k\pm}
    =0 .
\end{equation}
These are equivalent to the original complex fermion mode equation and
make explicit the imaginary contribution generated by the
time-dependent effective fermion mass.

\bibliography{refs.bib}

@article{Acharya:2001gy,
  author = {Acharya, Bobby S.},
  title = {On realising $N=1$ super Yang--Mills in M theory},
  journal = {hep-th},
  year = {2001},
  eprint = {hep-th/0011089}
}

@article{Acharya:2007rc,
  author = {Acharya, Bobby S. and Bobkov, Konstantin and Kane, Gordon and Kumar, Piyush and Shao, Jing},
  title = {Explaining the electroweak scale and stabilizing moduli in M theory},
  journal = {Phys. Rev. D},
  volume = {76},
  pages = {126010},
  year = {2007},
  eprint = {hep-th/0701034}
}

@article{Acharya:2008zi,
  author = {Acharya, Bobby S. and Bobkov, Konstantin and Kane, Gordon and Kumar, Piyush and Shao, Jing},
  title = {Moduli stabilization and the pattern of soft supersymmetry breaking terms},
  journal = {Phys. Rev. D},
  volume = {78},
  pages = {065038},
  year = {2008},
  eprint = {0801.0478}
}

@article{Giblin:2017wlo,
    author = "Giblin, John T. and Kane, Gordon and Nesbit, Eva and Watson, Scott and Zhao, Yue",
    title = "{Was the Universe Actually Radiation Dominated Prior to Nucleosynthesis?}",
    eprint = "1706.08536",
    archivePrefix = "arXiv",
    primaryClass = "hep-th",
    doi = "10.1103/PhysRevD.96.043525",
    journal = "Phys. Rev. D",
    volume = "96",
    number = "4",
    pages = "043525",
    year = "2017"
}

@article{Acharya:2010zx,
    author = "Acharya, Bobby Samir and Bobkov, Konstantin and Kumar, Piyush",
    title = "{An M Theory Solution to the Strong CP Problem and Constraints on the Axiverse}",
    eprint = "1004.5138",
    archivePrefix = "arXiv",
    primaryClass = "hep-th",
    doi = "10.1007/JHEP11(2010)105",
    journal = "JHEP",
    volume = "11",
    pages = "105",
    year = "2010"
}

@article{Acharya:2015zfk,
    author = "Acharya, Bobby Samir and Pongkitivanichkul, Chakrit",
    title = "{The Axiverse induced Dark Radiation Problem}",
    eprint = "1512.07907",
    archivePrefix = "arXiv",
    primaryClass = "hep-ph",
    doi = "10.1007/JHEP04(2016)009",
    journal = "JHEP",
    volume = "04",
    pages = "009",
    year = "2016"
}

@article{Greene:2000ew,
    author = "Greene, Patrick B. and Kofman, Lev",
    title = "{On the theory of fermionic preheating}",
    eprint = "hep-ph/0003018",
    archivePrefix = "arXiv",
    reportNumber = "CITA-2000-05",
    doi = "10.1103/PhysRevD.62.123516",
    journal = "Phys. Rev. D",
    volume = "62",
    pages = "123516",
    year = "2000"
}

@article{Coughlan1983,
  author  = {Coughlan, G. D. and Fischler, W. and Kolb, Edward W. and Raby, S. and Ross, G. G.},
  title   = {Cosmological Problems for the Polonyi Potential},
  journal = {Phys. Lett. B},
  volume  = {131},
  pages   = {59--64},
  year    = {1983},
  doi     = {10.1016/0370-2693(83)91091-2}
}

@article{deCarlos1993,
  author        = {de Carlos, B. and Casas, J. A. and Quevedo, F. and Roulet, E.},
  title         = {Model Independent Properties and Cosmological Implications of the Dilaton and Moduli Sectors of 4-D Strings},
  journal       = {Phys. Lett. B},
  volume        = {318},
  pages         = {447--456},
  year          = {1993},
  doi           = {10.1016/0370-2693(93)91538-X},
  eprint        = {hep-ph/9308325},
  archivePrefix = {arXiv}
}

@article{Banks1994,
  author        = {Banks, Tom and Kaplan, David B. and Nelson, Ann E.},
  title         = {Cosmological Implications of Dynamical Supersymmetry Breaking},
  journal       = {Phys. Rev. D},
  volume        = {49},
  pages         = {779--787},
  year          = {1994},
  doi           = {10.1103/PhysRevD.49.779},
  eprint        = {hep-ph/9308292},
  archivePrefix = {arXiv}
}

@article{Nelson_2018,
   title={Axion cosmology with early matter domination},
   volume={98},
   ISSN={2470-0029},
   url={http://dx.doi.org/10.1103/PhysRevD.98.063516},
   DOI={10.1103/physrevd.98.063516},
   number={6},
   journal={Physical Review D},
   publisher={American Physical Society (APS)},
   author={Nelson, Ann E. and Xiao, Huangyu},
   year={2018},
   month={Sept} }

@article{Amin_2012,
   title={Oscillons after Inflation},
   volume={108},
   ISSN={1079-7114},
   url={http://dx.doi.org/10.1103/PhysRevLett.108.241302},
   DOI={10.1103/physrevlett.108.241302},
   number={24},
   journal={Physical Review Letters},
   publisher={American Physical Society (APS)},
   author={Amin, Mustafa A. and Easther, Richard and Finkel, Hal and Flauger, Raphael and Hertzberg, Mark P.},
   year={2012},
   month={June} }

@article{DineRandallThomas1995,
  author        = {Dine, Michael and Randall, Lisa and Thomas, Scott},
  title         = {Supersymmetry Breaking in the Early Universe},
  journal       = {Phys. Rev. Lett.},
  volume        = {75},
  pages         = {398--401},
  year          = {1995},
  doi           = {10.1103/PhysRevLett.75.398},
  eprint        = {hep-ph/9503303},
  archivePrefix = {arXiv}
}

@article{Linde1996,
  author        = {Linde, Andrei},
  title         = {Relaxing the Cosmological Moduli Problem},
  journal       = {Phys. Rev. D},
  volume        = {53},
  pages         = {R4129--R4132},
  year          = {1996},
  doi           = {10.1103/PhysRevD.53.R4129},
  eprint        = {hep-th/9601083},
  archivePrefix = {arXiv}
}

@article{KaneSinhaWatson2015,
  author        = {Kane, Gordon and Sinha, Kuver and Watson, Scott},
  title         = {Cosmological Moduli and the Post-Inflationary Universe: A Critical Review},
  journal       = {Int. J. Mod. Phys. D},
  volume        = {24},
  pages         = {1530022},
  year          = {2015},
  doi           = {10.1142/S0218271815300220},
  eprint        = {1502.07746},
  archivePrefix = {arXiv},
  primaryClass  = {hep-ph}
}

@article{Acharya2006Hierarchy,
  author        = {Acharya, Bobby S. and Bobkov, Konstantin and Kane, Gordon L. and Kumar, Piyush and Vaman, Diana},
  title         = {An {M} Theory Solution to the Hierarchy Problem},
  journal       = {Phys. Rev. Lett.},
  volume        = {97},
  pages         = {191601},
  year          = {2006},
  doi           = {10.1103/PhysRevLett.97.191601},
  eprint        = {hep-th/0606262},
  archivePrefix = {arXiv}
}

@article{Acharya2007Moduli,
  author        = {Acharya, Bobby S. and Bobkov, Konstantin and Kane, Gordon L. and Kumar, Piyush and Shao, Jing},
  title         = {Explaining the Electroweak Scale and Stabilizing Moduli in {M} Theory},
  journal       = {Phys. Rev. D},
  volume        = {76},
  pages         = {126010},
  year          = {2007},
  doi           = {10.1103/PhysRevD.76.126010},
  eprint        = {hep-th/0701034},
  archivePrefix = {arXiv}
}

@article{Acharya2008G2MSSM,
  author        = {Acharya, Bobby S. and Bobkov, Konstantin and Kane, Gordon L. and Shao, Jing and Kumar, Piyush},
  title         = {The {$G_2$}-MSSM: An {M} Theory Motivated Model of Particle Physics},
  journal       = {Phys. Rev. D},
  volume        = {78},
  pages         = {065038},
  year          = {2008},
  doi           = {10.1103/PhysRevD.78.065038},
  eprint        = {0801.0478},
  archivePrefix = {arXiv},
  primaryClass  = {hep-ph}
}

@article{Acharya2008Nonthermal,
  author        = {Acharya, Bobby S. and Kumar, Piyush and Bobkov, Konstantin and Kane, Gordon and Shao, Jing and Watson, Scott},
  title         = {Non-thermal Dark Matter and the Moduli Problem in String Frameworks},
  journal       = {JHEP},
  volume        = {06},
  pages         = {064},
  year          = {2008},
  doi           = {10.1088/1126-6708/2008/06/064},
  eprint        = {0804.0863},
  archivePrefix = {arXiv},
  primaryClass  = {hep-ph}
}

@article{MoroiRandall2000,
  author        = {Moroi, Takeo and Randall, Lisa},
  title         = {Wino Cold Dark Matter from Anomaly-Mediated SUSY Breaking},
  journal       = {Nucl. Phys. B},
  volume        = {570},
  pages         = {455--472},
  year          = {2000},
  doi           = {10.1016/S0550-3213(99)00748-8},
  eprint        = {hep-ph/9906527},
  archivePrefix = {arXiv}
}

@article{Acharya2009WIMPMiracle,
  author        = {Acharya, Bobby S. and Kumar, Piyush and Kane, Gordon and Watson, Scott},
  title         = {A Non-thermal {WIMP} Miracle},
  journal       = {Phys. Rev. D},
  volume        = {80},
  pages         = {083529},
  year          = {2009},
  doi           = {10.1103/PhysRevD.80.083529},
  eprint        = {0908.2430},
  archivePrefix = {arXiv},
  primaryClass  = {hep-ph}
}

@article{Acharya2010Axiverse,
  author        = {Acharya, Bobby S. and Bobkov, Konstantin and Kumar, Piyush},
  title         = {An {M} Theory Solution to the Strong {CP} Problem and Constraints on the Axiverse},
  journal       = {JHEP},
  volume        = {11},
  pages         = {105},
  year          = {2010},
  doi           = {10.1007/JHEP11(2010)105},
  eprint        = {1004.5138},
  archivePrefix = {arXiv},
  primaryClass  = {hep-th}
}

@article{Acharya2012Generic,
  author        = {Acharya, Bobby S. and Kane, Gordon and Kumar, Piyush},
  title         = {Compactified String Theories---Generic Predictions for Particle Physics},
  journal       = {Int. J. Mod. Phys. A},
  volume        = {27},
  pages         = {1230012},
  year          = {2012},
  doi           = {10.1142/S0217751X12300128},
  eprint        = {1204.2795},
  archivePrefix = {arXiv},
  primaryClass  = {hep-ph}
}

@article{AcharyaPongkitivanichkul2016,
  author        = {Acharya, Bobby S. and Pongkitivanichkul, Chakrit},
  title         = {The Axiverse Induced Dark Radiation Problem},
  journal       = {JHEP},
  volume        = {04},
  pages         = {009},
  year          = {2016},
  doi           = {10.1007/JHEP04(2016)009},
  eprint        = {1512.07907},
  archivePrefix = {arXiv},
  primaryClass  = {hep-ph}
}

@article{KofmanLindeStarobinsky1997,
  author        = {Kofman, Lev and Linde, Andrei and Starobinsky, Alexei A.},
  title         = {Towards the Theory of Reheating After Inflation},
  journal       = {Phys. Rev. D},
  volume        = {56},
  pages         = {3258--3295},
  year          = {1997},
  doi           = {10.1103/PhysRevD.56.3258},
  eprint        = {hep-ph/9704452},
  archivePrefix = {arXiv}
}

@article{FelderEtAl2001,
  author        = {Felder, Gary and Garcia-Bellido, Juan and Greene, Patrick B. and Kofman, Lev and Linde, Andrei and Tkachev, Igor},
  title         = {Dynamics of Symmetry Breaking and Tachyonic Preheating},
  journal       = {Phys. Rev. Lett.},
  volume        = {87},
  pages         = {011601},
  year          = {2001},
  doi           = {10.1103/PhysRevLett.87.011601},
  eprint        = {hep-ph/0012142},
  archivePrefix = {arXiv}
}

@article{DeskinsGiblinCaldwell2013,
  author        = {Deskins, J. Tate and Giblin, John T., Jr. and Caldwell, Robert R.},
  title         = {Gauge Field Preheating at the End of Inflation},
  journal       = {Phys. Rev. D},
  volume        = {88},
  pages         = {063530},
  year          = {2013},
  doi           = {10.1103/PhysRevD.88.063530},
  eprint        = {1305.7226},
  archivePrefix = {arXiv},
  primaryClass  = {astro-ph.CO}
}

@article{DealEtAl2025,
  author        = {Deal, Robert Wiley and Barrowes, Leia and Giblin, John T., Jr. and Sinha, Kuver and Watson, Scott and Adams, Fred C.},
  title         = {Cosmological Moduli and Non-perturbative Production of Axions},
  journal       = {JHEP},
  volume        = {07},
  pages         = {154},
  year          = {2025},
  doi           = {10.1007/JHEP07(2025)154},
  eprint        = {2501.17229},
  archivePrefix = {arXiv},
  primaryClass  = {hep-ph}
}

@article{LeedomEtAl2025,
  author        = {Leedom, Jacob M. and Putti, Margherita and Righi, Nicole and Westphal, Alexander},
  title         = {Preheating Axions in String Cosmology},
  journal       = {JHEP},
  volume        = {04},
  pages         = {095},
  year          = {2025},
  doi           = {10.1007/JHEP04(2025)095},
  eprint        = {2411.18496},
  archivePrefix = {arXiv},
  primaryClass  = {hep-th}
}

@article{Dine:1995uk,
    author = "Dine, Michael and Randall, Lisa and Thomas, Scott D.",
    title = "{Supersymmetry Breaking in the Early Universe}",
    eprint = "hep-ph/9503303",
    archivePrefix = "arXiv",
    journal = "Phys. Rev. Lett.",
    volume = "75",
    pages = "398--401",
    year = "1995",
    doi = "10.1103/PhysRevLett.75.398"
}

@article{Banks:1995dt,
    author = "Banks, Tom and Berkooz, M. and Steinhardt, Paul J.",
    title = "{The Cosmological Moduli Problem, Supersymmetry Breaking,
              and Stability in Postinflationary Cosmology}",
    eprint = "hep-th/9501053",
    archivePrefix = "arXiv",
    journal = "Phys. Rev. D",
    volume = "52",
    pages = "705--716",
    year = "1995",
    doi = "10.1103/PhysRevD.52.705"
}

@article{deCarlos:1993jw,
    author = "de Carlos, B. and Casas, J. A. and Quevedo, F. and Roulet, E.",
    title = "{Model Independent Properties and Cosmological Implications of the Dilaton and Moduli Sectors of 4-d Strings}",
    eprint = "hep-ph/9308325",
    archivePrefix = "arXiv",
    journal = "Phys. Lett. B",
    volume = "318",
    pages = "447--456",
    year = "1993",
    doi = "10.1016/0370-2693(93)91538-X"
}

@article{CMB-S4:2026mge,
    author = "Trendafilova, Cynthia and others",
    collaboration = "CMB-S4",
    title = "{Sensitivity of Next-Generation CMB Surveys to Neutrinos and Other Light Relics}",
    eprint = "2608.07453",
    archivePrefix = "arXiv",
    primaryClass = "astro-ph.CO",
    reportNumber = "UT-WI-23-2026, FERMILAB-PUB-26-0443-PPD",
    month = "8",
    year = "2026",
    journal = "arXiv preprint"
    
}

\end{document}